\documentclass[12pt,a4paper]{article}
\usepackage{fullpage}
\usepackage[centertags]{amsmath}
\usepackage{amsbsy}
\usepackage{amsfonts}
\usepackage{amssymb}
\usepackage[dvips]{graphicx}
\usepackage[dvips,monochrome]{color}
\allowdisplaybreaks[1]
\newcommand{\lsim}{\mathrel{\mathop{\kern 0pt \rlap
  {\raise.2ex\hbox{$<$}}}
  \lower.9ex\hbox{\kern-.190em $\sim$}}}
\newcommand{\gsim}{\mathrel{\mathop{\kern 0pt \rlap
  {\raise.2ex\hbox{$>$}}}
  \lower.9ex\hbox{\kern-.190em $\sim$}}}

\newcommand{\keywords}[1]{{\bf Keywords:} \textsf{#1}}
\newcommand{\pacs}[1]{{\bf PACS:} \textsf{#1}}
\def\@fnsymbol#1{if case#1\hbox{}\or*\or\dagger\or\ddagger\or\mathcar''278\or\mathchar''27B\or|\or**\or\dagger\dagger\or\ddagger\ddagger\else\@ctrerr\fi\relax}
\long\def\symbolfootnote[#1]#2{\begingroup%
\def\thefootnote{\fnsymbol{footnote}}\footnote[#1]{#2}\endgroup}

\long\def\letterfootnote[#1]#2{\begingroup%
\def\thefootnote{\alph{footnote}}\footnote[#1]{#2}\endgroup}

\begin{document}

\begin{center}
\large\bfseries
Secondary Cosmic Ray particles due to GCR interactions \\ in the Earth's
atmosphere
\\[0.5cm]
\normalsize\normalfont

G. Battistoni$^1$, F. Cerutti$^2$, A. Fass\`o$^3$, A. Ferrari$^2$, 
M.V. Garzelli$^{1,\,}$\symbolfootnote[4]{$\,$ Corresponding author, {\it  
$\,$e-mail:} {\tt Maria.Garzelli@mi.infn.it} --
Invited talk presented at CSSP07, Carpathian 
Summer School of Physics, Exotic Nuclei \& Nuclear/Particle Astrophysics,
Sinaia, Romania, August 20 - 31 2007 }, 
\\
M.~Lantz$^4$, S.~Muraro$^1$, L.S. Pinsky$^5$, J. Ranft$^6$, S. Roesler$^2$, P.R. Sala$^1$

\vspace{0.3cm}
\small\itshape
$^{1}$University of Milano, Department of Physics, and INFN, Milano, Italy \\
$^{2}$CERN, Geneva, Switzerland \\
$^{3}$SLAC, Stanford, CA, US \\
$^{4}$Chalmers University, Department of Fundamental Physics, Goteborg, Sweden \\
$^{5}$University of Houston, Department of Physics, Houston, TX, US \\
$^{6}$Siegen University, Fachbereich 7 - Physik, Siegen, Germany \\
\end{center}

\begin{abstract}
Primary GCR interact with the Earth's atmosphere originating atmospheric
sho\-wers, thus giving rise to fluxes of secondary particles 
in the atmosphere. Electroma\-gnetic and hadronic interactions interplay 
in the production of these particles, whose detection is performed 
by means of complementary techniques in different energy ranges and 
at different depths in the atmosphere, down to the Earth's surface.

Monte Carlo codes are essential calculation tools which can describe
the complexity of the physics of these phenomena, thus allowing the
analysis of experimental data. 
However, these codes are affected by
important uncertainties, concerning, in particular, hadronic physics at
high energy.
In this paper we shall report some results concerning inclusive particle
fluxes and atmospheric shower properties as obtained using the FLUKA 
transport and interaction code. Some emphasis will also be given to the
validation
of the physics models of FLUKA involved in these calculations.
\\
\\ 
{\keywords GCR EAS, inclusive fluxes of secondary particles, 
Monte Carlo $\mathrm {\,\,\,\,\,\,\,\,\,\,\,\,\,\,\,\,\,\,\,\,\,\,\,\,\,\,\,\,\,\,\,\,\,\,\,\,\,}$ 
   models and codes
} 
\\
{\pacs 98.70.Sa, 13.85-t, 13.75-t}
\end{abstract}


\section{Introduction: EAS development and detection}

GCR interactions in the Earth's atmosphere, induced
by particles with energies high enough (E $\gsim 10^{14} - 10^{15}$ eV), 
may originate Extended Air Showers (EAS). 
Both the EM and the $\mu$ component of EAS  
can be detected and used  
to infer properties concerning GCR primary spectrum energy and
composition.
At present, 
a special
emphasis is put in the investigation of Very 
High and Ultra High Energy Cosmic Rays (VHECR and UHECR), 
through dedicated experiments
devoted to the detection of EAS originated by these particles,
due to the still open questions concerning the shape of the spectrum 
($2^{nd}$ knee, ankle), the composition ($\gamma$, $p$, heavy ions),
 and the origin of cosmic rays 
(galactic / extragalactic), their transport, acceleration and the GZK cut-off.

Information on GCR primary flux
and composition are inferred from the experimental observables by using
Monte Carlo (MC) simulations. In particular, while 
the fluo\-re\-scen\-ce technique, used
by many experiments, allows with some uncertainties to estimate 
the EM energy deposited in the atmosphere,
 the total primary energies can be obtained only 
after estimating the missing energy 
carried by other shower components ($\mu$, nucleons...),  
and this requires indeed 
MC simulations. 
Even the search for particular estimators, such as the S variable,
allows~to obtain primary
energy information in a way nearly independent of 
composition~\cite{rebel}, but
needs MC for calibrations.

In particular, in the AUGER experiment~\cite{auger} 
information on primary energies
are inferred 
by hybrid measurements and the fluorescence technique 
as far as the EM sector
is concerned (with an uncertainty amounting to $\sim$ 20\%), 
whereas the number of
$\mu$s at the Earth's surface, $N_\mu$, is still
related to total primary energies
by means of MC simulations. On the other hand, in the KASCADE and 
KASCADE-Grande experiments~\cite{kascade} 
information on primary energies are inferred from measures of $\mu$ and
$e$ with the aid of MC shower simulations. 
Both in AUGER and in KASCADE / KASCADE-Grande, 
as well as in many other EAS experiments, 
MCs are heavily needed 
to analyse data in terms of GCR mass composition.
At present there are however still important uncertainties, since
different MC models produce important differences on the interpretation of
the same data.
The critical ingredients in these MC codes are the physical models for
hadronic and nuclear interaction, and their implementation.
 
In this work we summarize some of the results which can be obtained by
MC simulation of EAS, giving examples taken from the use of the FLUKA code.

\section{MC simulation of EAS}

EAS simulations can be performed by means of MC codes, which include
hadronic and EM modules. This method leads, in a straightforward way,
to account for fluctuations in the evolution of different showers
initiated by primaries of the same type and energy.

The GCR primary spectrum covers a wide energy range, extending
on many orders of magnitude. This fact implies that it is very
difficult to have in a MC code a single hadronic
model able to describe primary and secondary hadron
interactions in the Earth's
atmosphere. Therefore in many cases different models, differing 
according to the energy range under study, have to be merged together.
 
At present the most diffused MC package for GCR induced shower simulation 
is CORSIKA~\cite{corsika}. 
As far as its hadronic sector is concerned, it  
distinguishes between high energy (E $\gsim$ 100 - 200 GeV) 
and low energy (E $\lsim$ 100 - 200 GeV) hadronic models. In the present
version, CORSIKA offers the choice, at high energy, among models like
QGSJET, QGSJET-II,
SIBYLL, DPMJET-2.55, NEXUS, EPOS, whereas, at low energy,
GHEISHA, UrQMD, FLUKA are available.
Other packages for GCR EAS simulation also exist. We just mention, among
the others, the AIRES~\cite{aires} and the Cosmos~\cite{cosmos} 
codes. 

In this paper, we focus instead on the performances 
of the FLUKA multipurpose code~\cite{flukacern}. 
FLUKA is a fully integrated, high precision, transport and interaction code, 
without the need of invoking other external packages.
Besides its use in the CORSIKA
package at energies $\lsim$~100 - 200 GeV, 
FLUKA has recently been used also as a stand\-alone code, 
complemented by the DPMJET-III code~\cite{dpmjet3}, 
for the simulation of EAS at energies $<$ $10^{16}$ eV. 
FLUKA simulations can, in principle, 
lead to results different from the ones made by FLUKA+CORSIKA, 
due to the implementation of different EM and hadronic models. 
FLUKA contains a highly accurate EM interaction model~\cite{MC2000em,isv}, 
not 
included in CORSIKA. 
As far as the hadronic sector is concerned, the following models
are available in FLUKA:\\ 
- low energy (E $<$ 20 TeV) h + A reactions can be simulated by means
of the PEANUT (PreEquilibrium Approach to NUclear Thermalization)
module, 
originally worked out for e\-ner\-gies $<$ tens of GeV~\cite{Trieste}, and 
recently improved and extended 
to higher e\-ner\-gies~\cite{VarennaAlfredo}.
This extension has superseded a previously existing DPM + INC model, working 
at~E~$>$~5~GeV.
\\ 
- high energy (E $>$ 20 TeV) h + A reactions can be described 
by means of the interface to the DPMJET-2.5 or III codes. \\ 
- A + A reactions can be simulated by means of an interface to the
RQMD2.4 code, written in Frankfurt and modified for insertion in
FLUKA~\cite{braz}, for E $<$ 5 GeV/A, and by means of the interface to the 
DPMJET-2.5 or III codes for E $>$ 5 GeV/A. Alternatively,
the superposition model is also available to roughly describe 
heavy-ion interactions. 

Constraints on theoretical models for particle interactions
can (or will) be obtained by data collected at
collider experiments, but, unfortunately, diffractive 
cross-sections 
are difficult to be measured
at colliders. Furthermore, present and near future accelerators 
do not and will not allow to test the physics of
particles at the highest GCR energies (E~$>$~1.5~10$^{18}$~eV, 
GCR spectrum tail).  
Thus, MC models used in the ana\-ly\-sis of EAS data are
still affected by important uncertainties, concerning in particular 
the hadronic sector.
Other constraints/checks on low-energy models,
besides the ones from accelerator data,
can come from astrophysical measurements of inclusive particle fluxes 
at different atmospheric depths, 
as performed by balloon-borne experiments and satellites.
We emphasize that 
for an accurate description of VHECR and UHECR EAS
also low-energy physics is important, e.g. the larger
is the distance from the shower core and the shower inclination,
the stronger is the required accuracy of the adopted low energy~models.

\subsection{Validation of the FLUKA models at low energies}

Validation of the FLUKA models at low energies has been
performed by means of com\-pa\-ri\-sons of their theoretical predictions 
with data collected at the accelerators. 
Data concerning inclusive particle fluxes in the atmosphere
down to the Earth's surface, have been then
used to assess the FLUKA performance on cosmic ray physics.

While the first issue can be addressed by using the version 
of the code officially di\-stri\-buted and available on the web, 
the second one requires the inclusion of further 
(geometrical, geomagnetical, astrophysical) assumptions and information. 
These elements have been inserted by means of the FLUKA user routines.

\paragraph{\bf Validation of FLUKA using accelerator data}

Data recently collected by the HARP and the
NA49 experiments have been nicely reproduced by means of FLUKA~\cite{isv}, 
Many others validations of the code, referred to previously published data, 
can be found in li\-te\-ra\-ture. 
Furthermore, data from the SPS collider, 
Tevatron and RHIC have been used to
successfully validate the DPMJET2.5 and III codes~\cite{ranft}.

\paragraph{\bf Validation of FLUKA using astrophysical data from
satellites and balloon-borne experiments}

\begin{figure}
\begin{center}
  \includegraphics[width=.24\textwidth]{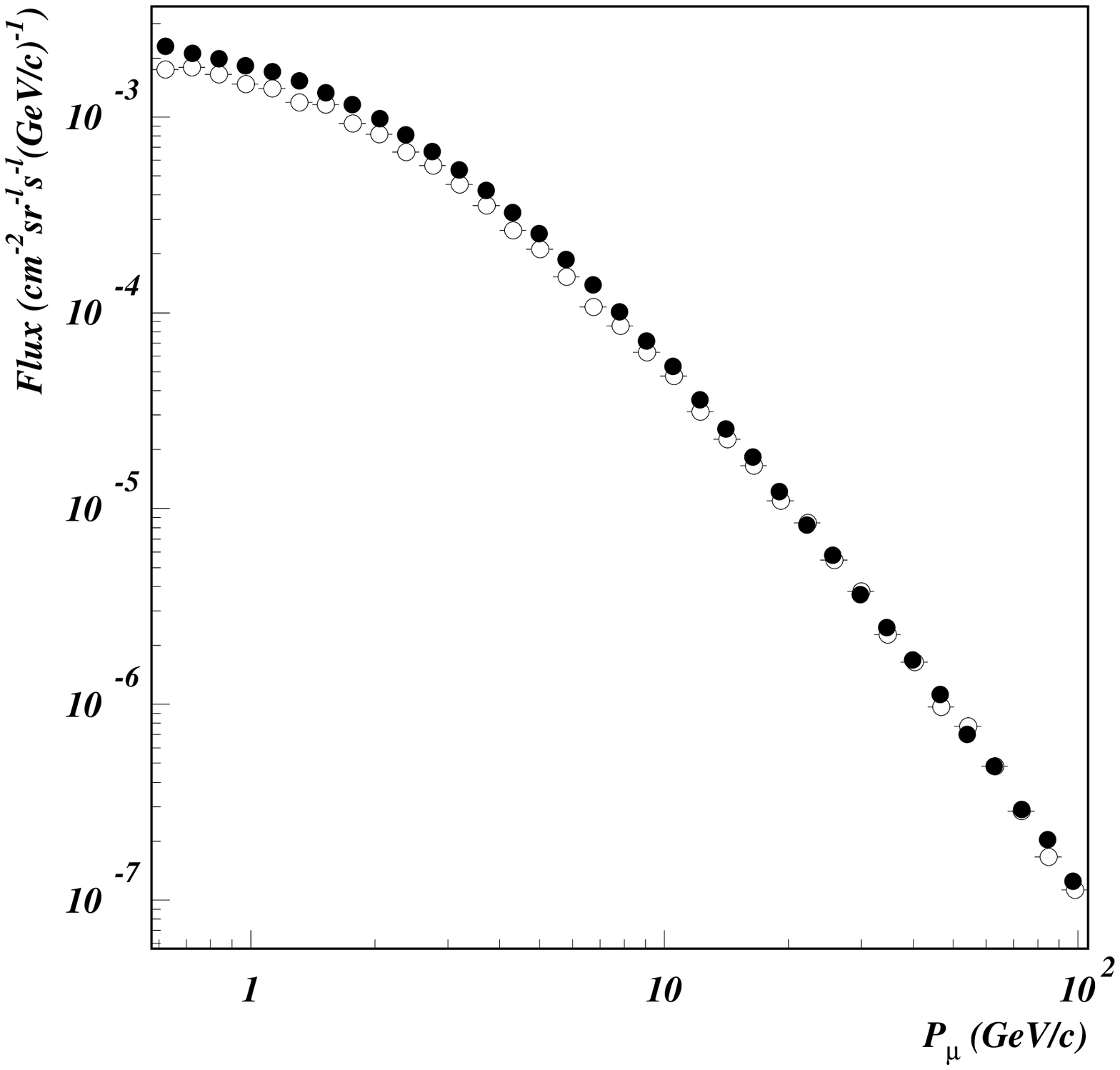}
  \includegraphics[width=.24\textwidth]{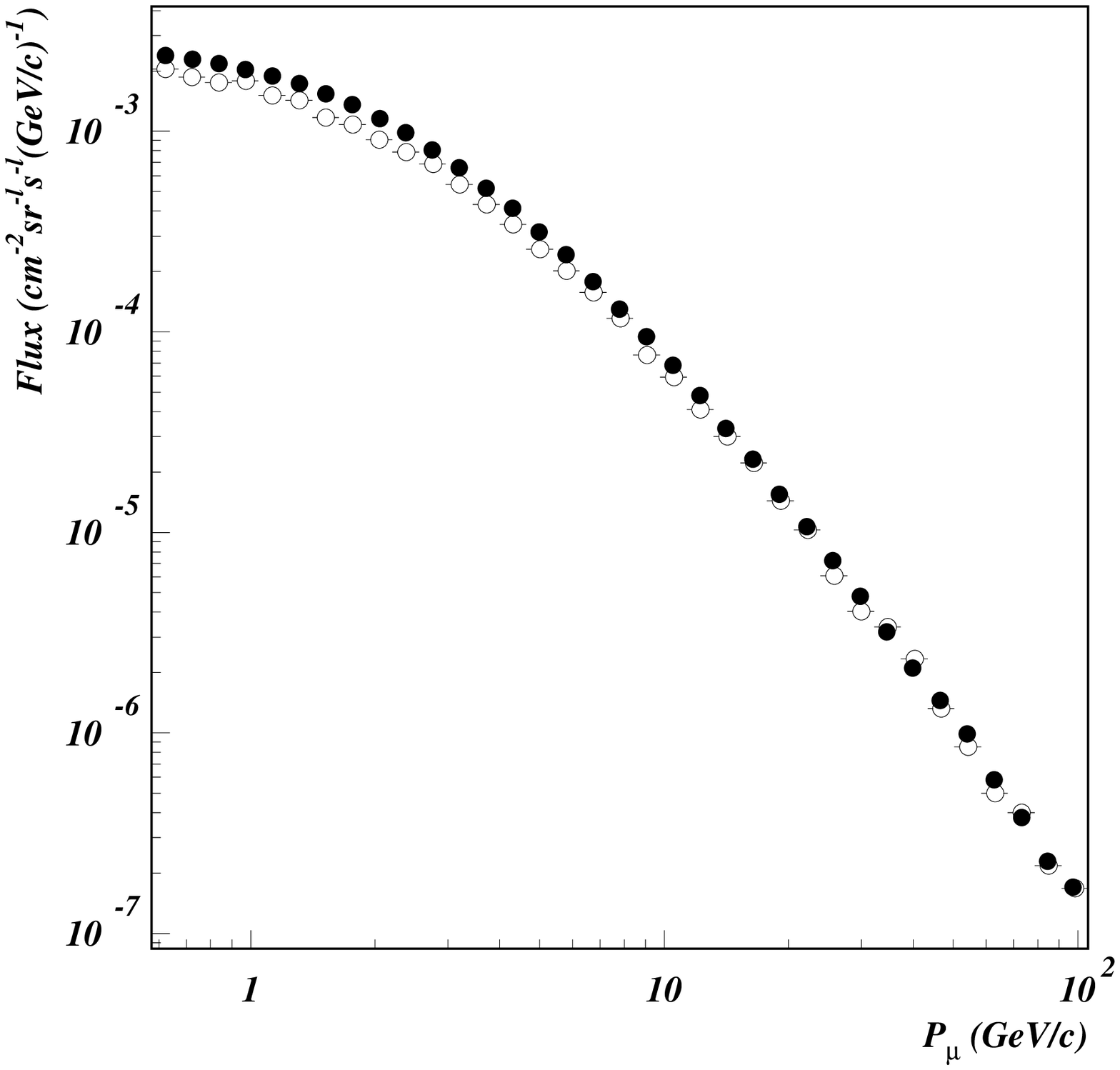}
  \caption{Calculated inclusive fluxes of $\mu^-$ ({\it left panel}) and
   $\mu^+$ ({\it right panel}) at the top of Mt. Norikura ($\sim$ 
   2700 m a.s.l.) as 
   obtained by FLUKA simulations (open symbols) vs. exp. 
   data~\cite{norikuraexp} from BESS'99 (full symbols).
Other examples of FLUKA benchmarks in GCR physics
can be found in~\cite{neutrini}.}  
   \label{norikura}
\end{center}
\end{figure}
\begin{figure}
\begin{center}
  \includegraphics[width=.24\textwidth]{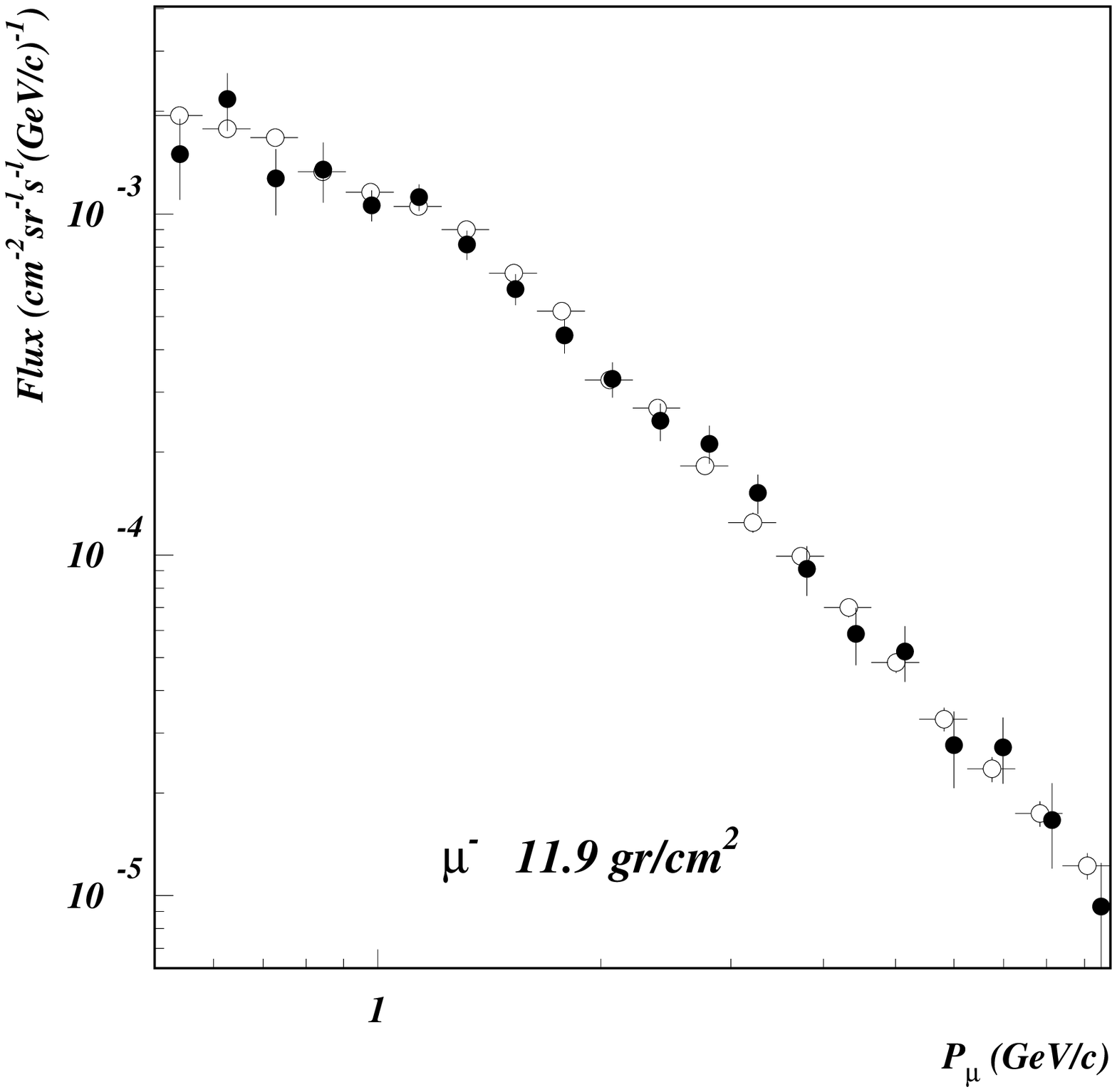}
  \includegraphics[width=.24\textwidth]{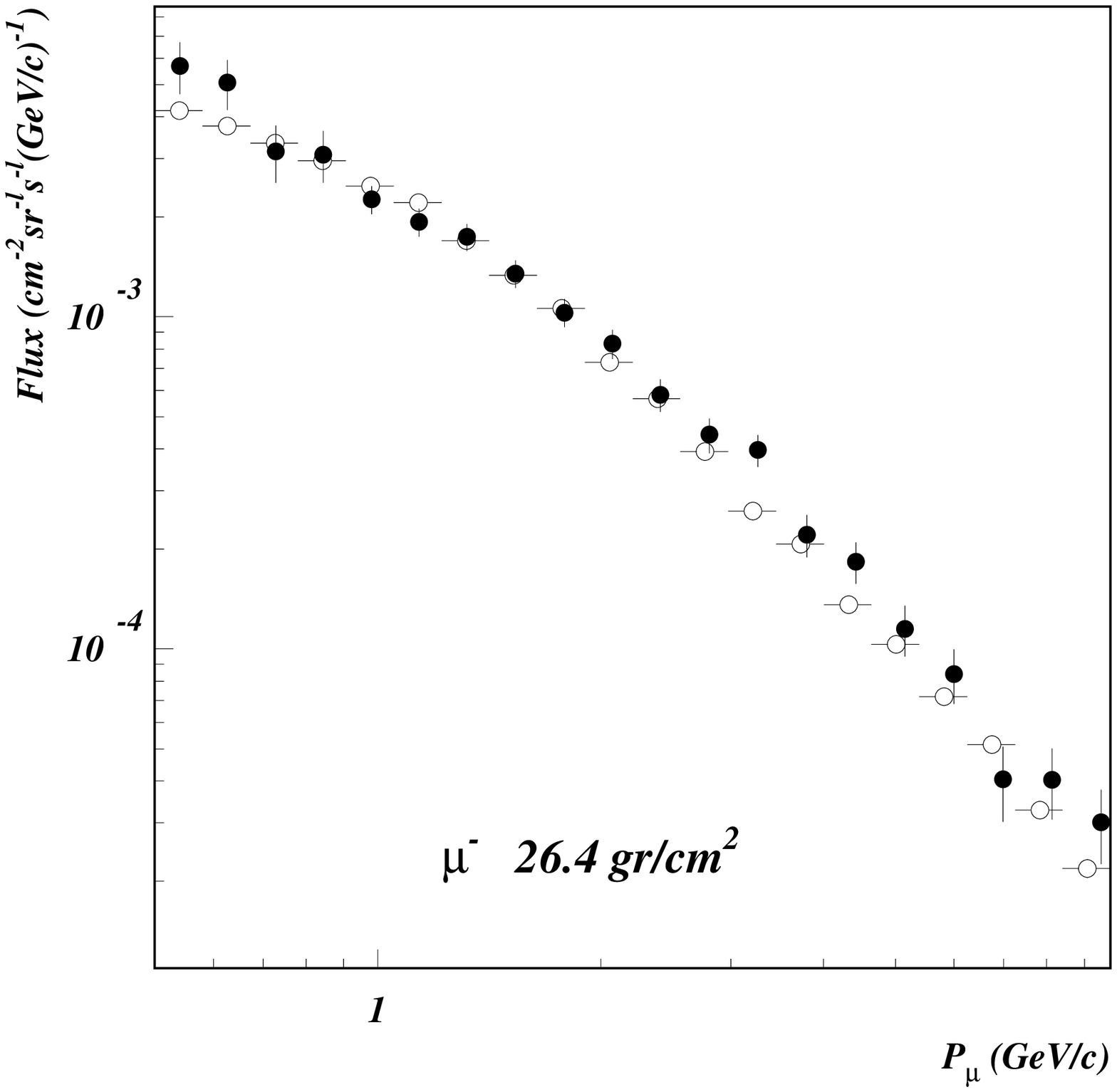}
  \includegraphics[width=.24\textwidth]{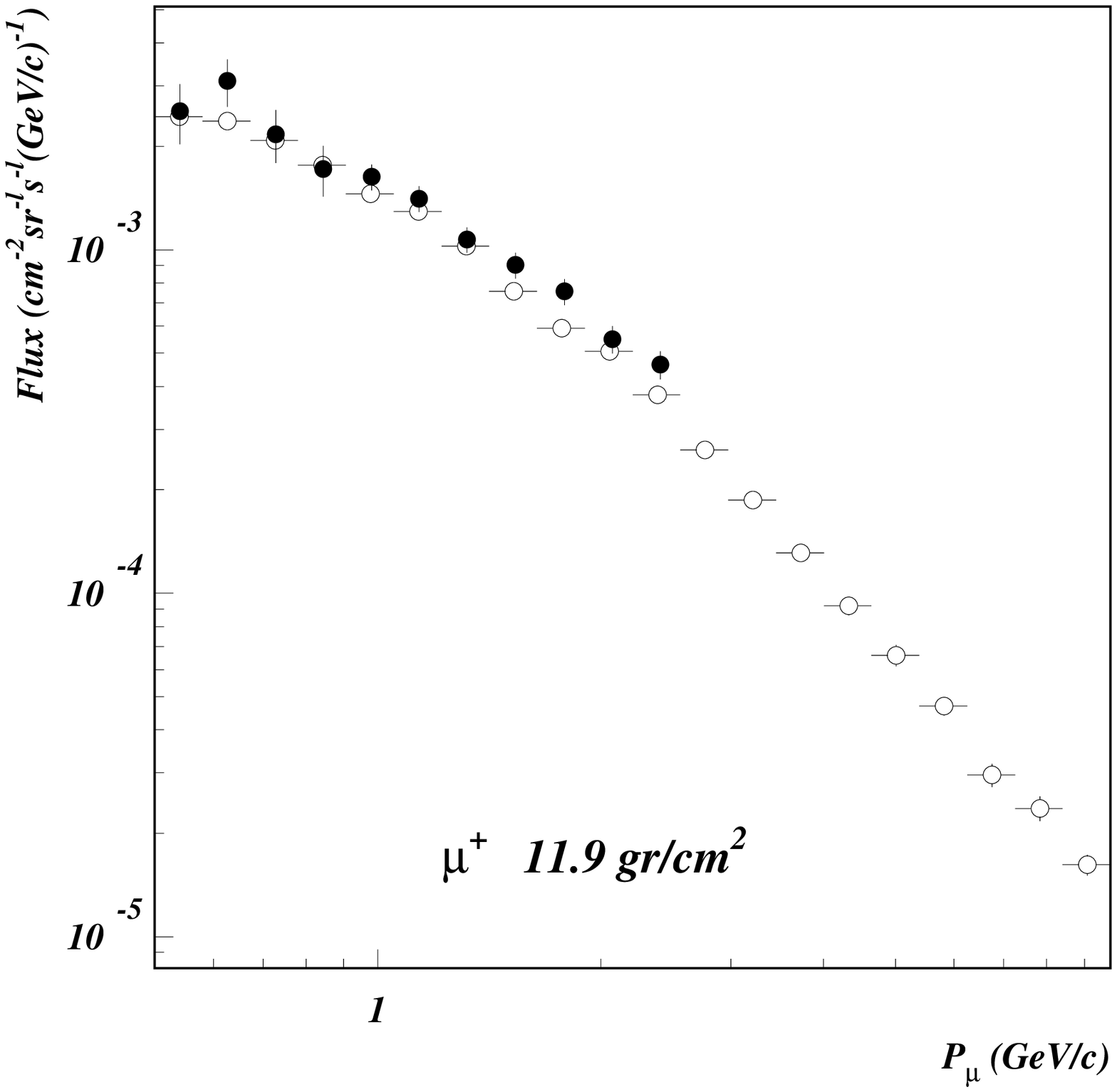}
  \includegraphics[width=.24\textwidth]{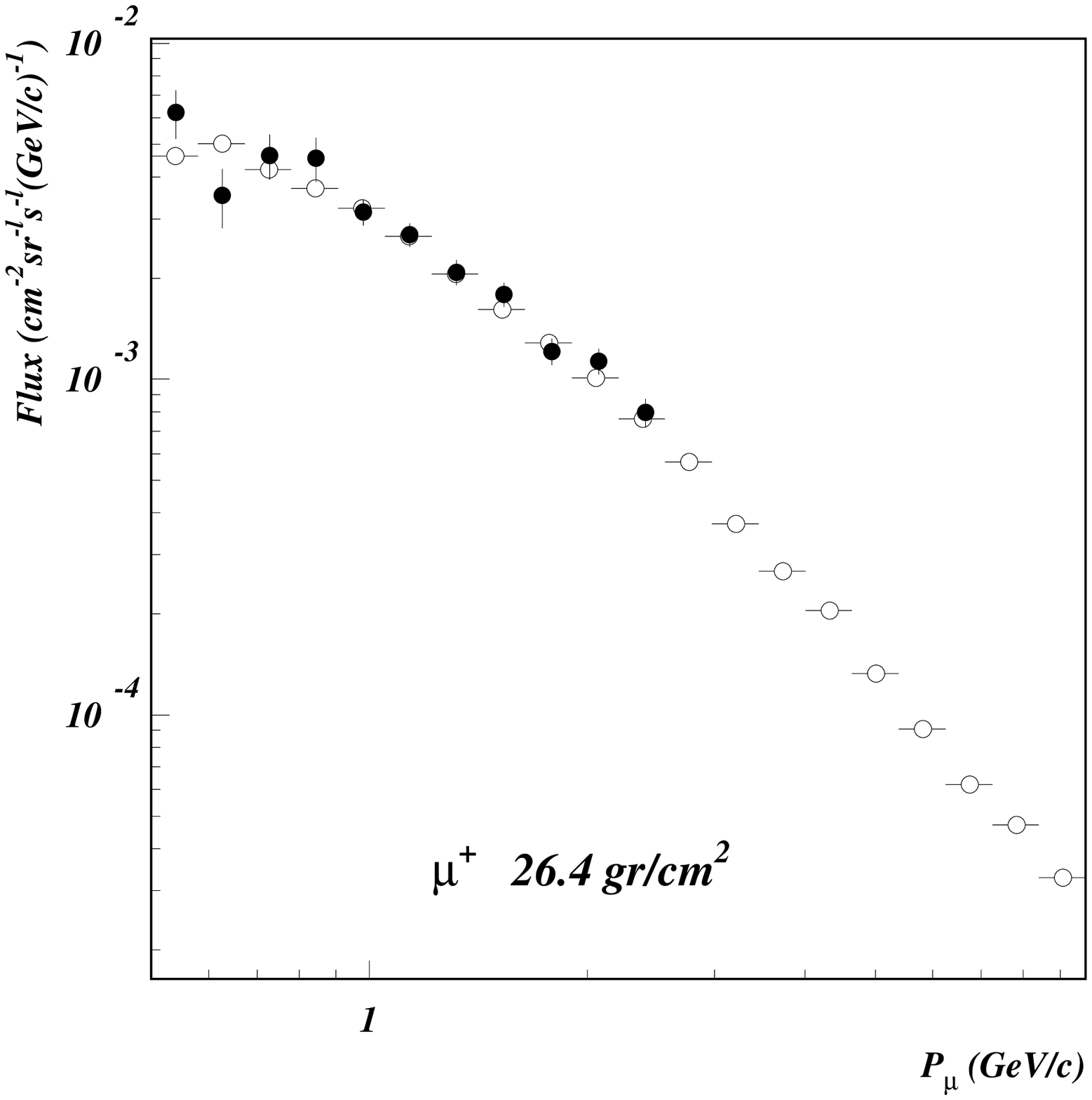}
  \caption{Calculated inclusive fluxes of $\mu^-$ ({\it left panels}) and
   $\mu^+$ ({\it right panels}) 
   at atmospheric depths of 11.9 g/cm$^2$ ({\it odd panels}) 
   and 26.4 g/cm$^2$ ({\it even panels}), respectively, as 
   obtained by FLUKA simulations (open symbols) vs. exp. 
   data~\cite{sumnerexp} from BESS 2001 (full symbols).}  
   \label{sumner}
\end{center}
\end{figure}

Inclusive particle fluxes in the Earth's upper and lower
atmosphere, induced by GCR propagation and interactions, 
have been measured by means of satellites and
balloon-borne experiments such as BESS. Different sets of
data, taken in dif\-fe\-rent atmospheric and geomagnetic conditions,
have been analyzed and reproduced by FLUKA, under proper
assumptions concerning GCR primary spectrum, solar modulation,
atmospheric and geomagnetic models. 
As examples, data concerning $\mu$ collected 
at the top  ($\sim$ 2700 m a.s.l.) of
Mt. Norikura (geo\-ma\-gne\-tic cut-off $\sim$ 11.2 GV) in 1999,
and at two different atmospheric depths (11.9 and
26.4 g/cm$^2$) in a balloon flight campaign
over Ft. Sumner (geomagnetic cut-off $\sim$ 4.2 GV)
in 2001, are presented in fig.~\ref{norikura} and ~\ref{sumner}, 
respectively, and compared to the theoretical fluxes expected
on the basis of FLUKA simulations. 
The agreement between the experimental and theoretical results 
is quite nice and increases at increasing energies. 
Furthermore, atmospheric $\mu$ charge ratios 
in the energy range from $\sim$ 20 GeV up to $\sim$ 3 TeV
have been detected by
the L3+C ex\-pe\-ri\-ment at CERN ($\sim$ 450 m a.s.l.), 
that has provided also $\mu$ arrival directions. 
FLUKA has successfully reproduced also these data~\cite{now2006}.

Proton and heavier ion inclusive fluxes 
have also been obtained by FLUKA theoretical simulations 
and compared to experimental data available from BESS. 
In these cases, the results of
the simulations turn out to be very sensitive to a correct
treatment of the geo\-ma\-gne\-tic cut-off, of the transport 
of particles in the upper layers of the atmosphere 
(trapping of particles and their recirculation),
and to the accuracy of the theoretical modelling of
low-energy A-A interactions.   
 

\subsection{Predictions of the FLUKA models at high energies: EAS}

In the following we shall report about some tests of EAS
simulation at energies 10$^{14}$ - 10$^{15}$~eV, 
performed by means of FLUKA, complemented by DPMJET-III
for the description of A-A reactions at E $>$ 5 GeV/A and h-A reactions
at E $>$ 20 TeV. 
Vertical showers, induced by primaries of different
mass ($p$ and Fe ions), have been considered.

\begin{figure}[t!]
  \begin{minipage}[t]{49mm}
  \includegraphics[bb=50 49 405 296, clip, width=1.\textwidth, height=4.8cm]{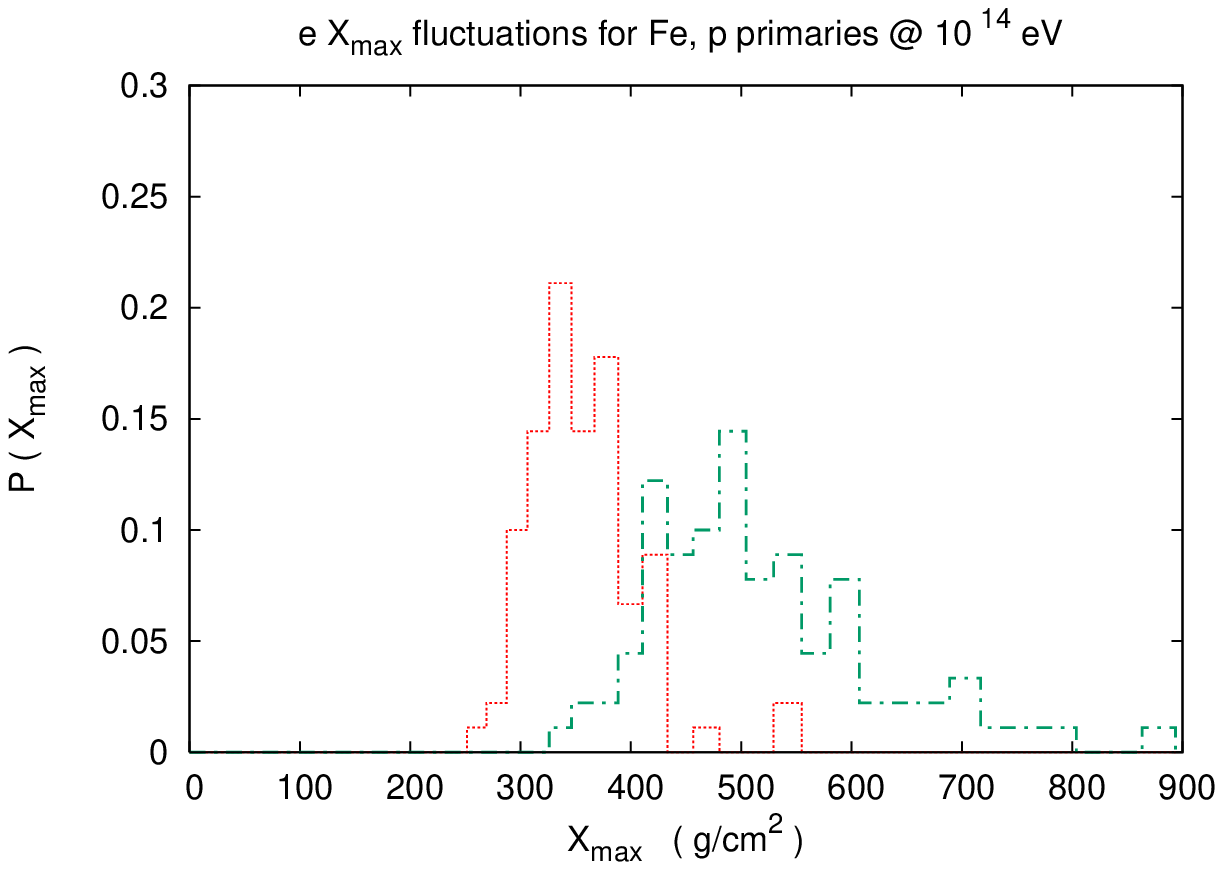}
  \end{minipage}
\hfill
  \begin{minipage}[t]{49mm}
\includegraphics[bb=50 49 405 296, clip, width=1.\textwidth]{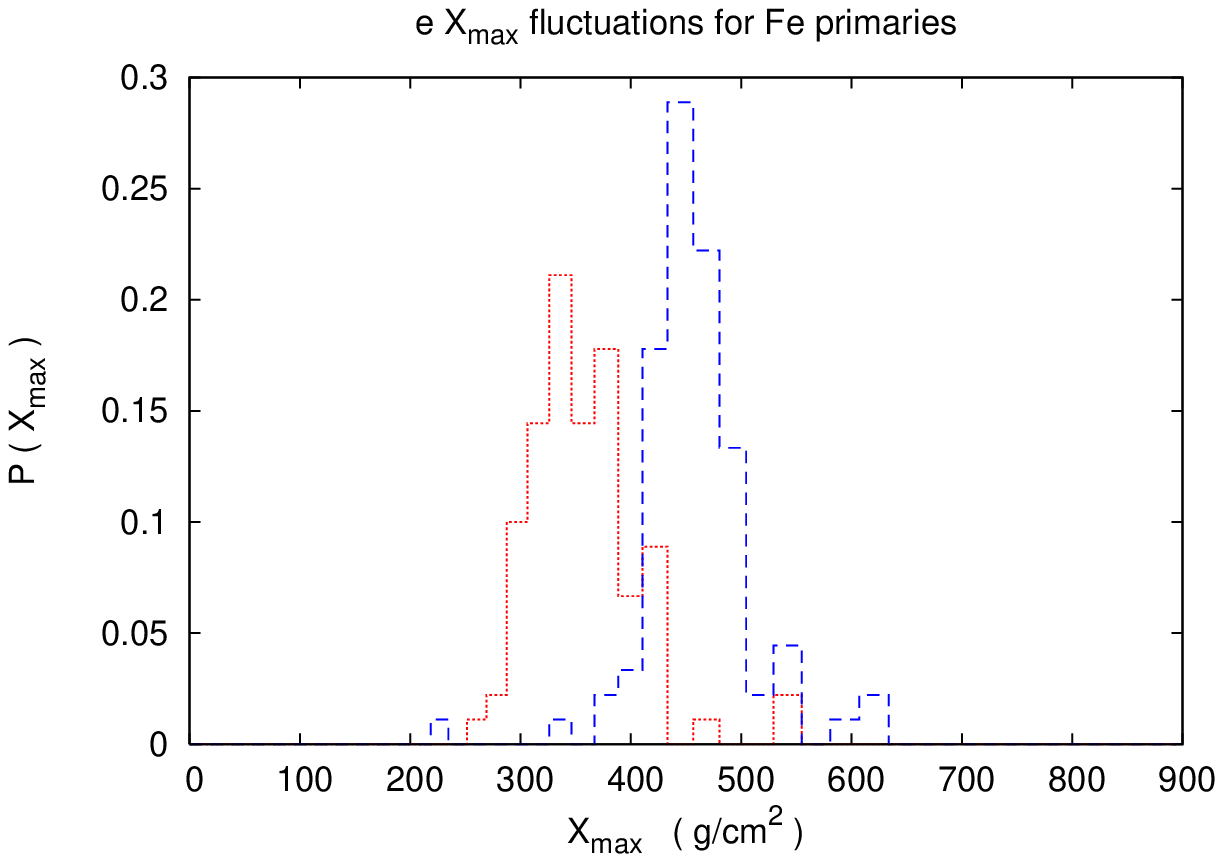}
\includegraphics[bb=50 49 405 296, clip, width=1.\textwidth]{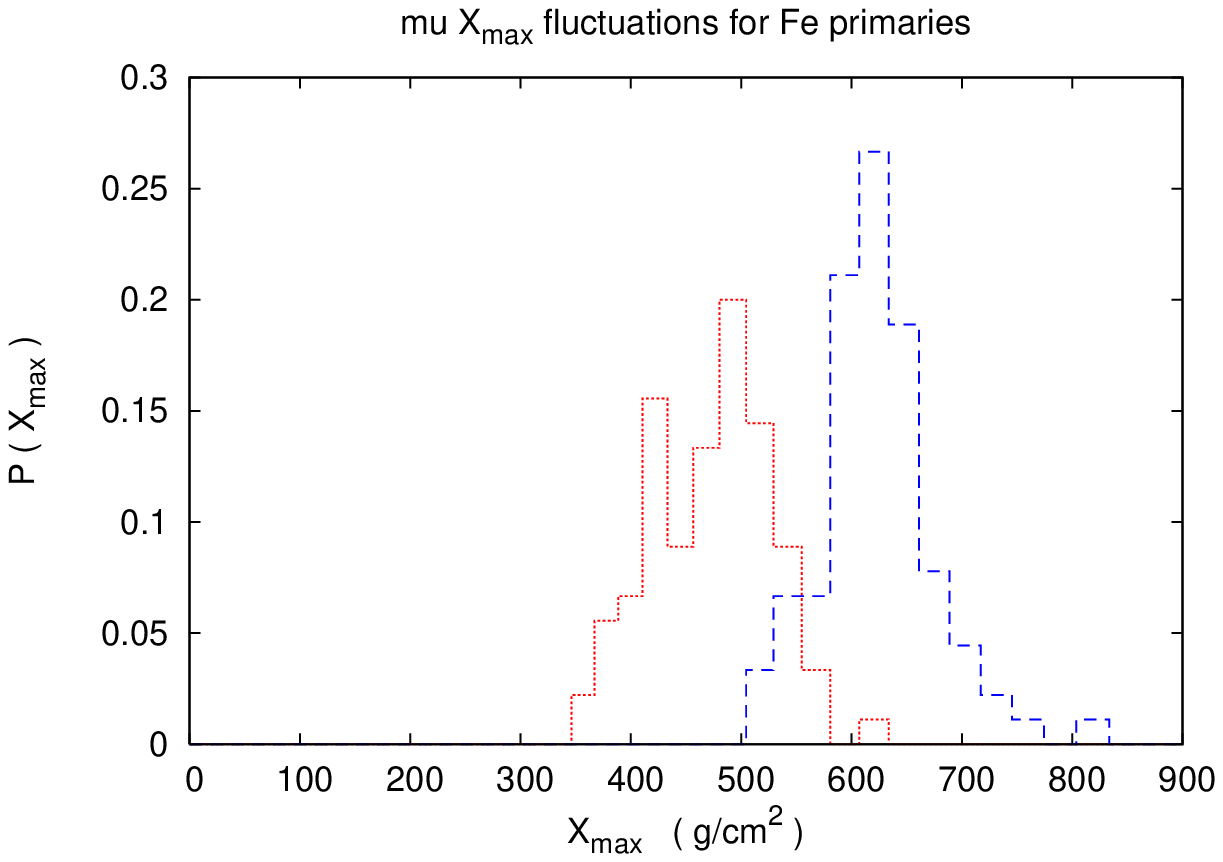}
  \end{minipage}
  \hfill
  \begin{minipage}[t]{49mm}
  \includegraphics[bb=50 49 405 296, clip, width=1.\textwidth]{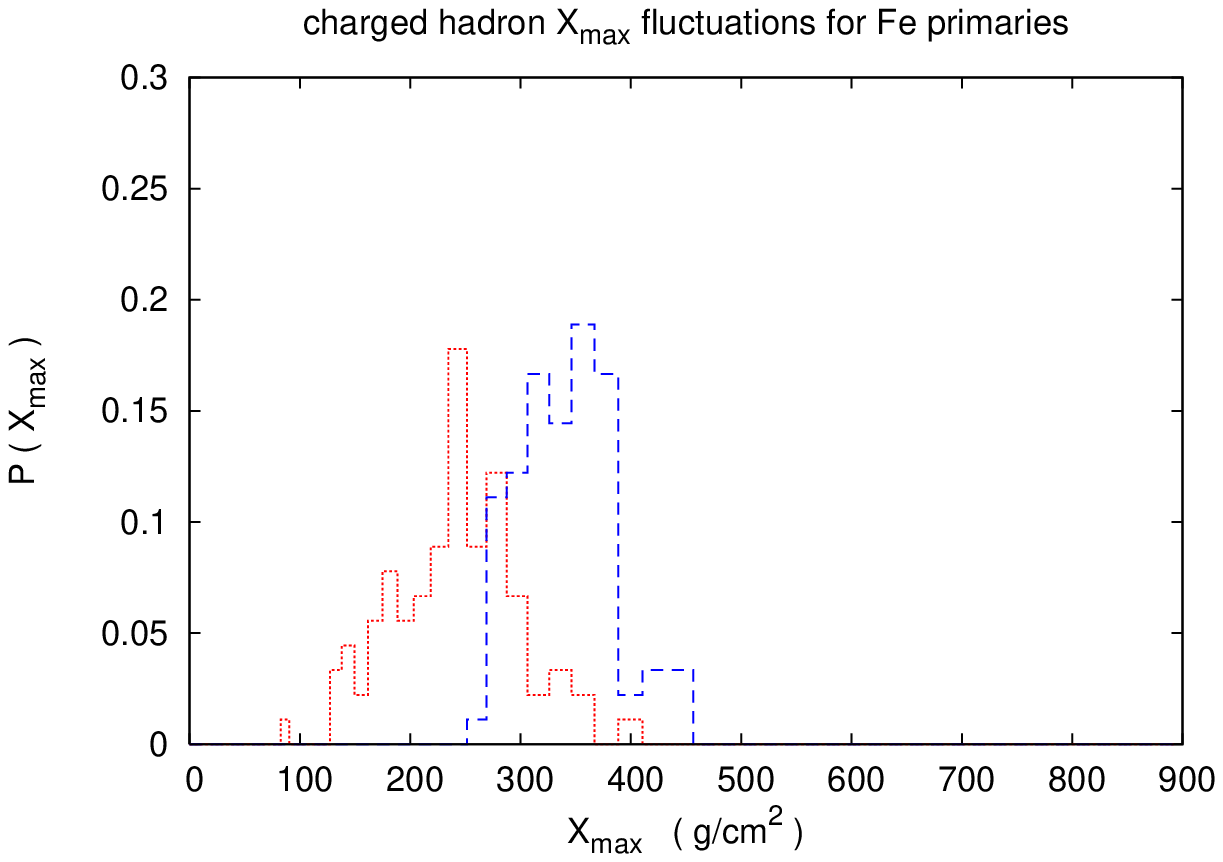}
  \includegraphics[bb=50 49 405 296, clip, width=1.\textwidth]{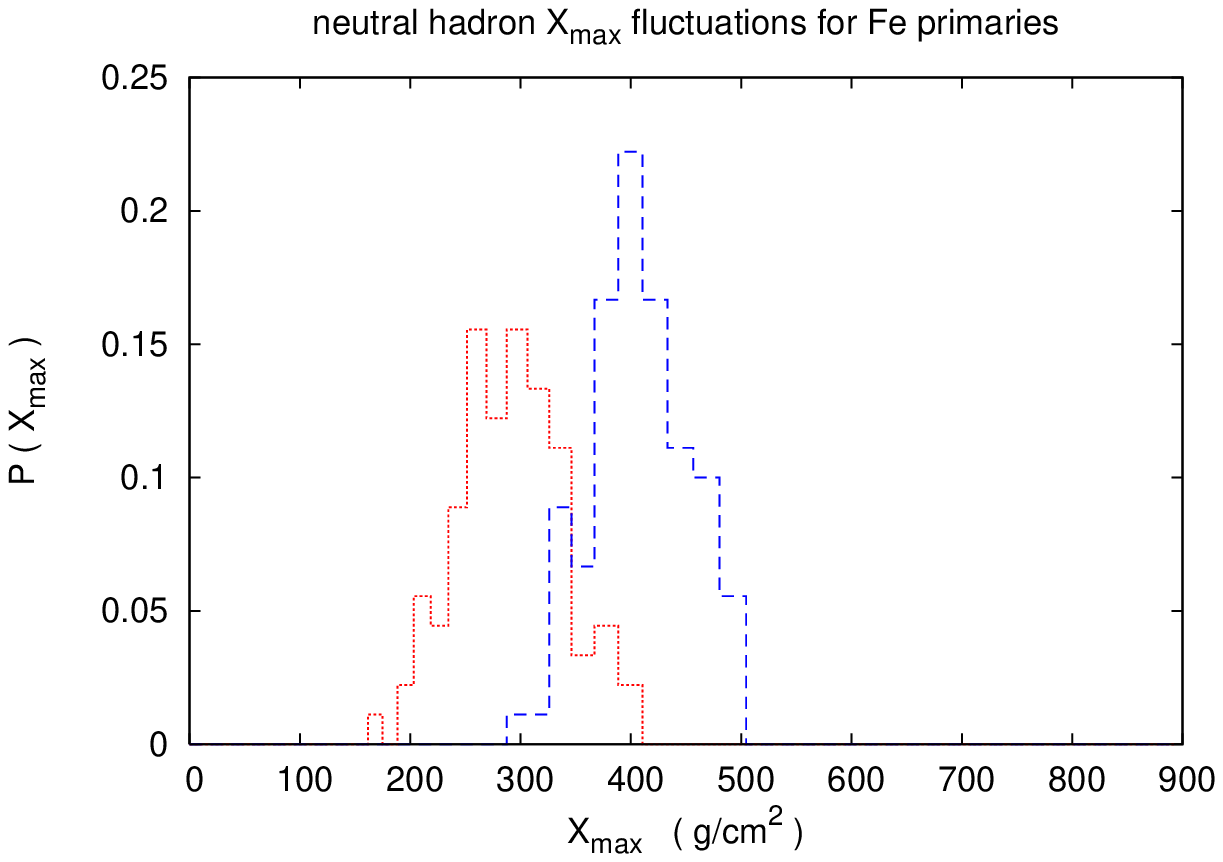}
  \end{minipage}
  \caption{ {\it Left panel}: 
   fluctuations in the distribution of $e$ $X_{\mathrm{max}}$ 
   for a few hundred  
   vertical showers induced by $p$ ({\it dot-dashed green line}) 
   and Fe ions ({\it dotted red line}) at 10$^{14}$ eV energy.
            {\it Central and Right panels}: 
     fluctuations in the distribution of $e$ $X_{\mathrm{max}}$ 
   ({\it central upper panel}), $\mu$ $X_{\mathrm{max}}$ ({\it central
    lower panel}), charged hadron $X_{\mathrm{max}}$
   ({\it right upper panel}) and neutral hadron $X_{\mathrm{max}}$ 
({\it right lower panel})
   for a few hundred  vertical showers 
   induced by Fe ions at 10$^{14}$ eV ({\it dotted red 
   line}) and 10$^{15}$ eV ({\it dashed blue line}) energies.}
   \label{fluelemu}
\end{figure}

The $X_{\mathrm{max}}$ position, i.e. the atmospheric depth where the
shower reaches its maximum development (in terms of EM energy deposition),  
is a good probe of GCR composition. In fact,
since the inelastic scattering cross-section for {\it p}-Air interactions is lower than the A-Air ones, 
EAS induced by Fe primaries have, on the average,
vertical profile
maxima shifted towards lower atmospheric depths 
than 
$p$ induced showers.
$e$ and $\mu$ fluences as a function of the atmospheric depth $\rho$ 
for $p$ and Fe induced showers at 10$^{14}$ and 10$^{15}$ eV,
as calculated with FLUKA + DPMJET-III, are shown in ref.~\cite{isv}.
In general, different codes, based on different models, 
due to different modelings of hadronic and EM processes,
give different results for $X_{\mathrm{max}}$. 
Anyway, to disentangle the effects of different models and different
primary composition, also more specific variables can be considered.
Among the others, the study of
$X_{\mathrm{max}}$ 
fluctuations~\cite{engel07} for primaries 
at a fixed energy has been proposed for this purpose.
As an example, $X_{\mathrm{max}}$ 
fluctuations for Fe induced showers
have been computed from our simulations, considering the $e$, $\mu$,
charged hadron and neutral hadron vertical shower profiles, 
and are shown in the central and right panels of fig.~\ref{fluelemu}. 
In general, the average muon $X_{\mathrm{max}}$ is located 
at larger $\rho$s with
respect to the electron one, since $\mu$s penetrate more
deeply. The charged hadrons undergo strong absorption effects,
thus their $X_{\mathrm{max}}$ are located at lower $\rho$s
with respect to the ones of neutral hadrons. Increasing
energies lead, in general, to a shift of $X_{\mathrm{max}}$ to higher 
$\rho$s for all considered profiles, even if,
in all cases, the fluctuation profiles at 10$^{14}$ and 
10$^{15}$ eV overlap. That means that reconstructing the energy 
of GCR primaries out of this method is not possible, since showers
of energies that differ even by a factor 10
can give rise, in some cases, to 
the same position of $X_{\mathrm{max}}$.
An example from our simulations
of the sensitivity of $X_{\mathrm{max}}$ fluctuations 
to the GCR primary composition is shown 
in the left panel of fig.~\ref{fluelemu}.
Not only 
the position of maxima but also the shape and broadness of the fluctuation
profiles 
change according to the GCR masses, as can be seen by comparing the 
cases of $p$ and $Fe$ induced EAS at a fixed energy (10$^{14}$ eV for the
cases considered in the figure). 

Besides the vertical spreads of particle distribution, also
lateral spreads can be calculated.
In particular, in case of EM particles, the lateral spread is due 
to the effect of multiple Coulomb scattering, 
while, in case of $\mu$s, the
lateral spread is mainly correlated with the $p_T$ distribution
of charged $\pi$ and $K$, from the decay of whom $\mu$s 
originate (we are not mentioning here $\mu$s from the decay of short-lived
charmed particles, which are also considered in FLUKA and other codes).  
Detailed maps of $e$ and $\mu$ fluences in the atmosphere, down 
to the Earth's surface, which allow to appreciate both the average 
vertical profile and the lateral spread of each component, 
as calculated by FLUKA for Fe induced
showers, can be found in~\cite{isv}.

\IfFileExists{\jobname.bbl}{}
 {\typeout{}
  \typeout{******************************************}
  \typeout{** Please run "bibtex \jobname" to optain}
  \typeout{** the bibliography and then re-run LaTeX}
  \typeout{** twice to fix the references!}
  \typeout{******************************************}
  \typeout{}
 }


\end{document}